\documentclass[sigconf,authorversion]{acmart}
\makeatletter
\renewcommand\@formatdoi[1]{\ignorespaces}
\makeatother
\usepackage[noend]{algorithmic}
\usepackage{algorithm}
\usepackage{multirow}
\usepackage{flushend}
\usepackage{amsmath,amssymb}
\usepackage{mathtools}
\usepackage[acronym]{glossaries}
\usepackage{units}
\usepackage{booktabs}
\usepackage[inline]{enumitem} 

\DeclareMathOperator*{\argmax}{arg\,max}

\setcopyright{none}

\acmConference[REVEAL '19]{the ACM RecSys Workshop on Reinforcement Learning and Robust Estimators for Recommendation Systems}{September 20th, 2019}{Copenhagen, Denmark}
\acmYear{2019}
\copyrightyear{2019}

\begin{document}
\title[Learning from Bandit Feedback]{Learning from Bandit Feedback:\\An Overview of the State-of-the-art}

\author{Olivier Jeunen}
\affiliation{
  \institution{University of Antwerp}
  \city{Antwerp}
  \country{Belgium}
}
\email{olivier.jeunen@uantwerp.be}

\author{Dmytro Mykhaylov}
\affiliation{
  \institution{Criteo AI Lab}
  \city{Paris}
  \country{France}
}
\email{d.mykhaylov@criteo.com}

\author{David Rohde}
\affiliation{
  \institution{Criteo AI Lab}
  \city{Paris}
  \country{France}
}
\email{d.rohde@criteo.com}

\author{Flavian Vasile}
\affiliation{
  \institution{Criteo AI Lab}
  \city{Paris}
  \country{France}
}
\email{f.vasile@criteo.com}

\author{Alexandre Gilotte}
\affiliation{
  \institution{Criteo AI Lab}
  \city{Paris}
  \country{France}
}
\email{a.gilotte@criteo.com}

\author{Martin Bompaire}
\affiliation{
  \institution{Criteo AI Lab}
  \city{Paris}
  \country{France}
}
\email{m.bompaire@criteo.com}

\renewcommand{\shortauthors}{O. Jeunen et al.}

\begin{abstract}
In machine learning we often try to optimise a decision rule that would have worked well over a historical dataset; this is the so called empirical risk minimisation principle.
In the context of learning from recommender system logs, applying this principle becomes a problem because we do not have available the reward of decisions we did not do.
In order to handle this ``bandit-feedback'' setting, several Counterfactual Risk Minimisation (CRM) methods have been proposed in recent years, that attempt to estimate the performance of different policies on historical data.
Through importance sampling and various variance reduction techniques, these methods allow more robust learning and inference than classical approaches.
It is difficult to accurately estimate the performance of policies that frequently perform actions that were infrequently done in the past and a number of different types of estimators have been proposed.

In this paper, we review several methods, based on different off-policy estimators, for learning from bandit feedback.
We discuss key differences and commonalities among existing approaches, and compare their empirical performance on the RecoGym simulation environment.
To the best of our knowledge, this work is the first comparison study for bandit algorithms in a recommender system setting.
\end{abstract}

%
%

\maketitle

\section{Learning from Bandit Feedback}
Traditional approaches to recommendation are often based on some form of collaborative filtering on the user-item matrix containing \emph{organic} user-item interactions~\cite{Hu2008,Rendle2008,Christakopoulou2016,Liang2018}.
These methods find their origin in the broader field of supervised learning, and generally don't take \emph{bandit} feedback into account (i. e. which recommendations were actually shown and whether the user interacted with them).
Recently, several methods have been proposed for so-called Batch Learning from Bandit-Feedback (BLBF) or Counterfactual Risk Minimisation (CRM)~\cite{Swaminathan2015JMLR}.
These methods make use of action-reward pairs: recommendations that were shown and whether they were interacted with.
Using counterfactual estimators~\cite{Owen2013}, they aim to learn an optimal recommendation policy.
This line of research is more closely related to the reinforcement learning field than classical supervised learning approaches~\cite{Sutton1998}.

In this work, we present several of those recently proposed methods and discuss their practicality in a recommender system context.
Throughout a series of experiments with the RecoGym simulation environment~\cite{Rohde2018}, we compare their empirical performance under various environments.

\subsection{Notations}
Throughout this work, we will denote the user state or context vector by $\mathbf{x} \in \mathbb{R}^n$.
Although this vector can be of arbitrary dimension, we will assume it to be a vector of length $n$ containing counts of historical interactions with items for fair comparison and simplicity.
An action is either represented by a scalar identifier $a$, or a one-hot encoded vector $\mathbf{a}$.
The reward for a given action, i. e. whether it leaded to a click, is denoted by $c \in \{0,1\}$.
In what follows, we provide an overview of the methods we discuss in this work, along with a brief explanation.
All models were optimised through the full-batch L-BFGS algorithm.
Table~\ref{tab:methods} provides an overview of the methods we include in our comparison.

\subsection{Methods}
\begin{table}[t]
    \centering
    \begin{tabular}{lccccc}
        \toprule
        \textbf{Method} &    $P(c|x,a)$ &   $P(a|x)$   &   \textbf{IPS} &   \textbf{Variance}    \\
        \midrule
        Likelihood &   \checkmark  &   &   &   &\\
        IPS Likelihood &   \checkmark  &   &  \checkmark &   &\\
        Contextual Bandit &   &  \checkmark   & \checkmark  &   &\\
        Dual Bandit &  \checkmark  &  \checkmark   & \checkmark  &   &\\
        POEM &    &  \checkmark   & \checkmark  & \checkmark  &\\
        SNIPS &    &  \checkmark   & \checkmark  & \checkmark  &\\
        \bottomrule
    \end{tabular}
    \caption{An overview of the methods we discuss in this paper, and how they relate to one another.}
    \label{tab:methods}
\end{table}
\paragraph{Likelihood.}
The standard statistical approach is to do statistical inference and decision making in two separate steps.  The simplest approach is to first do statistical inference using maximum likelihood and then do decision making in a separate step, bypassing the empirical/counterfactual risk minimisation approach.  If the reward is a binary variable, then a logistic regression model is appropriate:

\begin{equation}\label{eq:likelihood}
    P(c|\mathbf{x},a) = \sigma((\mathbf{x}\otimes\mathbf{a})^\intercal\mathbf{\beta})
\end{equation}

Here, $\mathbf{\beta} \in \mathbb{R}^{n^2}$ are the model parameters and $\sigma(\cdot)$ is the logistic sigmoid; $\otimes$ is the Kronecker product.

\paragraph{Reweighted Likelihood.}

Often we fit a simple model e.g. a linear model to a complex relationship.  When we do this, the model will underfit the data and is unable to capture the true relationship.  When a standard maximum likelihood approach is used the error due to the underfitting will be minimized around common occurrences of $(\mathbf{x},a)$.  This leads to a phenomenon widely known as \emph{covariate shift}~\cite{Shimodaira2000}.
One general solution to this issue is to make use of importance sampling~\cite{Owen2013}, and reweight samples to adjust for the difference in the distribution of past actions (as per the logging policy) and future actions (which we will evaluate uniformly in this case).
Practically, this is achieved by reweighting samples $(\mathbf{x}_i,a_i,c_i)$ by the inverse propensity score of the logging policy during maximum likelihood estimation: $w_i = \frac{1}{\pi_0(a_i|\mathbf{x}_i)}$.  This procedure is only beneficial if the model lacks capacity to correctly model the complete relationship \cite{storkey}.

\paragraph{Contextual Bandits (CB)}
The previous approaches model the probability of a click, given a context-action pair.
Contextual bandits aim to directly model the probability of an action, given a context vector~\cite{Langford2008}.
This is shown in Equation~\ref{eq:contextual}, where $\theta \in \mathbb{R}^{n\times n}$ are the model parameters.
\begin{equation}\label{eq:contextual}
    P(a|\mathbf{x}) = \text{softmax}(\mathbf{x}^\intercal\mathbf{\theta})
\end{equation}

The goal at hand is to learn a policy that chooses the optimal action given a context $\mathbf{x}$, i.e. the policy that maximises the number of clicks we would have gotten when $\pi_{\theta}$ was deployed instead of the logging policy $\pi_0$.
Equation~\ref{eq:counterfactual_objective} formalises this counterfactual objective.
\begin{equation}\label{eq:counterfactual_objective}
    \theta^{*} = \argmax_{\theta}\sum_{i = 1}^{N}c_i \frac{\pi_\theta(a_i,\mathbf{x_i})}{\pi_0(a_i,\mathbf{x_i})}
\end{equation}

We optimise it by maximising a lower bound on the log of the objective obtained through Jensen's inequality~\cite{Jensen1906}, since it then becomes the log-likelihood of a multinomial logistic regression where each observation has been weighted by $w_i = \frac{c_i}{\pi_0(a_i|\mathbf{x}_i)}$.
In doing so, it improves numerical stability of the optimisation procedure.

\paragraph{Dual Bandit.}
Due to the reweighting scheme used in the contextual bandit approach, the model ends up learning only from those context-action pairs that led to a click.
Indeed, if $c_i = 0$ then $w_i = 0$.
Nevertheless, valuable information is embedded in the negative samples about what actions should \emph{not} be repeated in the future.
One possible approach is to jointly optimise the contextual bandit objective with the likelihood presented in Equation~\ref{eq:likelihood}.
Here, $\beta$ is a flattened version of $\theta$, and the model parameters are shared.
This leads to the combined loss shown in Equation~\ref{eq:dual_loss}, where $0 \leq \alpha \leq 1$ is a hyper-parameter that controls the influence of the negative log-likelihood on the final loss.
\begin{equation}\label{eq:dual_loss}
    \mathcal{L}_{\text{dual}}(\theta) = (1 - \alpha) \mathcal{L}_{\text{CB}}(\theta) + \alpha \mathcal{L}_{\text{LH}}(\theta)
\end{equation}

\paragraph{Policy Optimiser for Exponential Models (POEM)}
Inverse propensity scoring is a powerful technique that allows for counterfactual optimisation.
When the target policy $\pi_\theta$ and the logging policy $\pi_0$ diverge, however, IPS-based estimators tend to fail as the variance of the estimate grows along.
\citeauthor{Swaminathan2015} tackle this by formalising the Counterfactual Risk Minimisation (CRM) principle along with a learning algorithm POEM~\cite{Swaminathan2015}.
On top of the standard counterfactual objective presented in Equation~\ref{eq:counterfactual_objective}, CRM clips the IPS weights~\cite{Ionides2008} and includes a variance penalisation term.
We do not clip importance weights in our experiments.

\paragraph{Self-normalised IPS (SNIPS)}
Variance regularisation alone, as included in POEM, is insufficient to fully avoid overfitting in the bandit-feedback case.
Norm-POEM is an extension of POEM that optimises the CRM objective for the SNIPS estimator~\cite{Swaminathan2015snips}, specifically targeting the problem of propensity overfitting.
This work is further extended and generalised to allow for optimisation through stochastic gradient descent (SGD) methods~\cite{Joachims2018}.

\section{Results and Conclusion}
\begin{figure}
    \centering
    \includegraphics[scale = .40, trim = {4mm 10mm 0cm 0cm}]{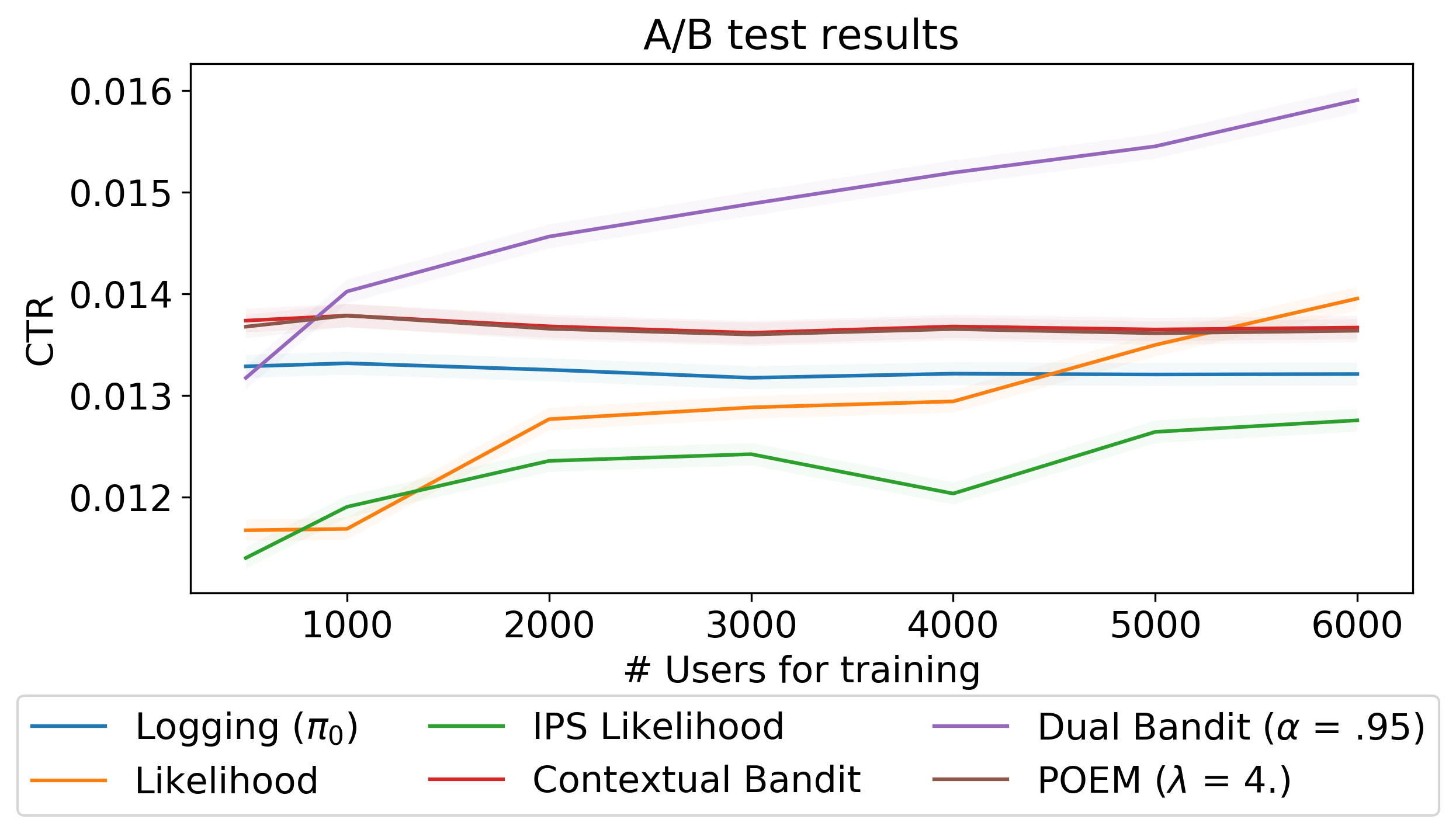}
    \caption{Simulated A/B test results}
    \label{fig:results}
\end{figure}

Figure~\ref{fig:results} shows the measured click-through-rate (CTR) for a varying number of users in the training data, a fixed set of $10$ items, and a reasonable popularity-based logging policy\footnote{All code necessary to replicate our experiments will be open-sourced upon acceptance.}.
For $5\,000$ users, we approximately obtain $400\,000$ logged bandit events.
We evaluate every method on $30\,000$ users, in order to obtain a robust estimate of performance.
Around every line, the 95\% confidence interval is shown.
We see the contextual bandit approach being the most effective when not much data is available, and the variance penalisation scheme from POEM not having a big impact.
This latter observation might change drastically for a less sensible logging policy or a larger number of actions, as this would impose more uncertainty.
Likelihood catches up when fed more training data, and the combined objective from the dual bandit stands out, showing merit in including negative feedback for policy learning.

We have presented several recently proposed methods for learning from bandit feedback, and discussed their practicality in a recommender system context.
Through experimental validation on the RecoGym environment, we empirically validated the performance of said approaches.
As future work, we aim to include more promising recent methods such as the one presented in~\cite{su2019cab}, and perform more experiments on the impact of the logging policy on the quality of the learning procedures.

\bibliographystyle{ACM-Reference-Format}
\bibliography{bibliography}

\end{document}